\documentclass[12pt,nohyper]{JHEP3}
\usepackage{graphics,psfrag}
\usepackage{amsthm,amssymb,epsfig,amsmath,euscript,array,cite}

\newcounter{multieqs}

\newcommand{\be}{\begin{equation}}
\newcommand{\ee}{\end{equation}}
\newcommand{\eq}[1]{(\ref{#1})}

\newcommand{\bm}[1]{\mbox{\boldmath $#1$}}
\newcommand{\rf}[1]{(\ref{#1})}

\def\bd{\begin{document}}
\def\ed{\end{document}}
\def\nn{\nonumber}
\def\bea{\begin{eqnarray}}
\def\eea{\end{eqnarray}}
\let\bm=\bibitem
\let\la=\label

\def\npb#1#2#3{Nucl. Phys. {\bf{B#1}} #3 (#2)}
\def\plb#1#2#3{Phys. Lett. {\bf{#1B}} #3 (#2)}
\def\prl#1#2#3{Phys. Rev. Lett. {\bf{#1}} #3 (#2)}
\def\prd#1#2#3{Phys. Rev. {D \bf{#1}} #3 (#2)}
\def\cmp#1#2#3{Comm. Math. Phys. {\bf{#1}} #3 (#2)}
\def\cqg#1#2#3{Class. Quantum Grav. {\bf{#1}} #3 (#2)}
\def\nppsa#1#2#3{Nucl. Phys. B (Proc. Suppl.) {\bf{#1A}}#3 (#2)}
\def\ap#1#2#3{Ann. of Phys. {\bf{#1}} #3 (#2)}
\def\ijmp#1#2#3{Int. J. Mod. Phys. {\bf{A#1}} #3 (#2)}
\def\rmp#1#2#3{Rev. Mod. Phys. {\bf{#1}} #3 (#2)}
\def\mpla#1#2#3{Mod. Phys. Lett. {\bf A#1} #3 (#2)}
\def\jhep#1#2#3{J. High Energy Phys. {\bf #1} #3 (#2)}
\def\atmp#1#2#3{Adv. Theor. Math. Phys. {\bf #1} #3 (#2)}

\def\N{{\cal N}}
\def\sst{\scriptscriptstyle}
\def\thetabar{\bar\theta}
\def\Tr{{\rm Tr}}
\def\one{\mbox{1 \kern-.59em {\rm l}}}

\def\a{\alpha}      \def\da{{\dot\alpha}}  \def\dA{{\dot A}}
\def\b{\beta}       \def\db{{\dot\beta}}  
\def\g{\gamma}  \def\G{\Gamma}  \def\dc{{\dot\gamma}}  
\def\d{\delta}  \def\D{\Delta}  \def\ddt{\dot\delta}  
\def\e{\epsilon}        \def\ve{\varepsilon}  
\def\f{\phi}    \def\F{\Phi}    \def\vvf{\f}  
\def\h{\eta}  
\def\k{\kappa}  
\def\l{\lambda} \def\L{\Lambda}  
\def\m{\mu} \def\n{\nu}  
\def\o{\omega}  
\def\p{\pi} \def\P{\Pi}  
\def\r{\rho}  
\def\s{\sigma}  \def\S{\Sigma}  
\def\t{\tau}  
\def\th{\theta} \def\Th{\Theta} \def\vth{\vartheta}  
\def\X{\Xeta}  
\def\z{\zeta}  

\def\na{\nabla}  
\def\cA{{\cal A}} \def\cB{{\cal B}} \def\cC{{\cal C}}  
\def\cD{{\cal D}} \def\cE{{\cal E}} \def\cF{{\cal F}}  
\def\cG{{\cal G}} \def\cH{{\cal H}} \def\cI{{\cal I}}  
\def\cJ{{\cal J}} \def\cK{{\cal K}} \def\cL{{\cal L}}  
\def\cM{{\cal M}} \def\cN{{\cal N}} \def\cO{{\cal O}}  
\def\cP{{\cal P}} \def\cQ{{\cal Q}} \def\cR{{\cal R}}  
\def\cS{{\cal S}} \def\cT{{\cal T}} \def\cU{{\cal U}}  
\def\cV{{\cal V}} \def\cW{{\cal W}} \def\cX{{\cal X}}  
\def\cY{{\cal Y}} \def\cZ{{\cal Z}}

\def\ua{\underline{\alpha}}  
\def\uc{\underline{\phantom{\alpha}}\!\!\!\gamma}  
\def\um{\underline{\mu}}  
\def\ud{\underline\delta}  
\def\ue{\underline\epsilon}  
\def\una{\underline a}\def\unA{\underline A}  
\def\unb{\underline b}\def\unB{\underline B}  
\def\unc{\underline c}\def\unC{\underline C}  
\def\und{\underline d}\def\unD{\underline D}  
\def\une{\underline e}\def\unE{\underline E}  
\def\unf{\underline{\phantom{e}}\!\!\!\! f}\def\unF{\underline F}  
\def\unm{\underline m}\def\unM{\underline M}  
\def\unn{\underline n}\def\unN{\underline N}  
\def\unp{\underline{\phantom{a}}\!\!\! p}\def\unP{\underline P}  
\def\unq{\underline{\phantom{a}}\!\!\! q}  
\def\unQ{\underline{\phantom{A}}\!\!\!\! Q}  
\def\unH{\underline{H}}  
  
\def\As {{A \hspace{-6.4pt} \slash}\;}  
\def\bs {{b \hspace{-6.4pt} \slash}\;}  
\def\Ds {{D \hspace{-6.4pt} \slash}\;}  
\def\ds {{\del \hspace{-6.4pt} \slash}\;}  
\def\ss {{\s \hspace{-6.4pt} \slash}\;}  
\def\ks {{ k \hspace{-6.4pt} \slash}\;}  
\def\ps {{p \hspace{-6.4pt} \slash}\;}   
\def\xs {{x \hspace{-6.4pt} \slash}\;}  
\def\pas {{{p_1} \hspace{-6.4pt} \slash}\;}  
\def\pbs {{{p_2} \hspace{-6.4pt} \slash}\;}   
\def\cFs {{{\cal F} \hspace{-6.4pt} \slash}\;}

\def\Dh{\hat{D}}
\def\Gh{\hat{G}}
\def\Fh{\hat{F}}
\def\Ph{\hat{P}}
\def\Rh{\hat{R}}
\def\Vh{\hat{V}}  
\def\Xh{\hat{X}} 
 
\def\ah{\hat{a}}
\def\gh{\hat{g}} 
\def\hh{\hat{h}}
\def\uh{\hat{u}}  
\def\xh{\hat{x}}  
\def\yh{\hat{y}}  
\def\ph{\hat{p}}  
\def\xih{\hat{\xi}}  
\def\chih{\hat{\chi}}

\def\psit{\tilde{\psi}}  
\def\Psit{\tilde{\Psi}}   
\def\Psibt{\tilde{\bar{Psi}}}  

\def\st{\tilde{\sigma}}  

\def\Phit{\tilde{\Phi}}   
\def\Phitb{\overline{\tilde{Phi}}}  
\def\tht{\tilde{\th}}  
\def\lt{\tilde{\l}}
\def\chit{\tilde{\chi}}   
\def\phit{\tilde{\phi}} 

\def\At{\tilde{A}}
\def\Bt{\tilde{B}}
\def\Ct{\tilde{C}}
\def\Dt{\tilde{D}}
\def\Ft{\tilde{F}}
\def\Qt{\tilde{Q}}  
\def\Rt{\tilde{R}}  
\def\Mt{\tilde{M }}  
\def\Nt{\tilde{N}}   
\def\St{\tilde{S}}
\def\Vt{\tilde{V}}
\def\Xt{\tilde{X}} 
\def\at{\tilde{a}}
\def\ct{\tilde{c}}   
\def\htt{\tilde{h}} 
\def\ft{\tilde{f}}
\def\gt{\tilde{g}}
\def\pt{\tilde{p}}  
\def\qt{\tilde{q}}  
\def\vt{\tilde{v}}  
\def\nt{\tilde{n}}  
\def\ut{\tilde{u}}  
\def\wt{\tilde{w}}  
\def\zt{\tilde{z}} 
\def\xt{\tilde{x}} 
\def\yt{\tilde{y}} 
\def\Psit{\tilde{\Psi}}
\def\vphit{\tilde{\varphi}}  

\def\delb{\bar{\partial}}  
\def\thb{\bar{\theta}}
\def\mub{\bar{\mu}}
\def\lamb{\bar{\l}}
\def\psib{\bar{\psi}}
\def\sb{\bar{\sigma}}
\def\xib{\bar{\xi}}
\def\chib{\bar{\chi}}

\def\Phib{\bar{\Phi}}
\def\Lamb{\bar{\Lambda}}
\def\Sb{{\overline \Sigma}}
\def\cb{\bar{c}}
\def\wb{\bar{w}}
\def\ub{\bar{u}}
\def\zb{{\bar{z}}}
\def\Qb{{\bar Q}}
\def\qb{\bar{q}}

\def\Ab{{\overline A}} \def\Bb{{\overline B}} \def\Cb{{\overline C}}  
\def\Db{{\overline D}} \def\Eb{{\overline E}} \def\Fb{{\overline F}}  
\def\Gb{{\overline G}} \def\Hb{{\overline H}} \def\Ib{{\overline I}}  
\def\Jb{{\overline J}} \def\Kb{{\overline K}} \def\Lb{{\overline L}}  
\def\Mb{{\overline M}} \def\Nb{{\overline N}} \def\Ob{{\overline O}}  
\def\Pb{{\overline P}}  \def\Rb{{\overline R}}  
 \def\Tb{{\overline T}} \def\Ub{{\overline U}}  
\def\Vb{{\overline V}} \def\Wb{{\overline W}} \def\Xb{{\overline X}}  
\def\Yb{{\overline Y}} \def\Zb{{\overline Z}}  

\def\fb{{\overline f}}
\def\gb{{\overline g}}
\def\mb{{\overline m}}
\def\lb{{\overline l}}
\def\yb{{\overline y}}

\def\ba{{\bf a}} 
\def\bk{{\bf k}}  
\def\bl{{\bf l}}  
\def\bp{{\bf p}}  
\def\bq{{\bf q}}  
\def\br{{\bf r}}
\def\bt{{\bf t}}
\def\bu{{\bf u}}
\def\bv{{\bf v}}
\def\bx{{\bf x}}  
\def\by{{\bf y}}  
\def\bR{{\bf R}}  
\def\bV{{\bf V}}

\def\bone{{\bf 1}}  

\def\va{{\vec a}}
\def\vk{{\vec k}}
\def\vp{{\vec p}}
\def\vq{{\vec q}}
\def\vx{{\vec x}}
\def\vy{{\vec y}}
\def\vu{{\vec u}}
\def\vv{{\vec v}}

\def\vs{{\vec \sigma}}
\def\vtau{{\vec \tau}}

\newcommand{\ov}[1]{\overrightarrow{#1}}
  
\def\d{\delta}\def\D{\Delta}\def\ddt{\dot\delta}  
  
\def\pa{\partial} \def\del{\partial}  
\def\xx{\times}  
\def\uno{\mbox{1 \kern-.59em {\rm l}}}    
  
\def\trp{^{\top}}  
\def\inv{^{-1}}  
\def\dag{{^{\dagger}}}  
\def\pr{^{\prime}}  
  
\def\rar{\rightarrow}  
\def\lar{\leftarrow}  
\def\lrar{\leftrightarrow}  
  
\newcommand{\0}{\,\!}      %this is just NOTHING!  
\def\one{1\!\!1\,\,}  
\def\im{\imath}  
\def\jm{\jmath}  
  
\newcommand{\tr}{\mbox{tr}}  
\newcommand{\slsh}[1]{/ \!\!\!\! #1}  
  
\def\vac{|0\rangle}  
\def\lvac{\langle 0|}  
  
\def\hlf{\frac{1}{2}}  
\def\ove#1{\frac{1}{#1}}  

\def\Box{\square}  
\def\CC {\mathbb{C}}
\def\RR{\mathbb{R}}
\def\ZZ{\mathbb{Z}}  
\def\bb#1{{\bf #1}}  
\def\bcomment#1{}  
\def\bfhat#1{{\bf \hat{#1}}}  
\def\VEV#1{\left\langle #1\right\rangle}  

\newcommand{\ex}[1]{{\rm e}^{#1}} \def\ii{{\rm i}}  

\newcommand{\lrbrk}[1]{\left(#1\right)}
\newcommand{\sfrac}[2]{{\textstyle\frac{#1}{#2}}}
 
\def\stw{{\sqrt{2}}}

\def\rf {{\rm f}}
\def\ri {{\rm i}}
\def\rs {{\scriptscriptstyle \rm S}}
\def\rt {{\scriptscriptstyle \rm T}}

\def\rQ {{\scriptscriptstyle \rm \cQ}}
\def\rR {{\scriptscriptstyle \rm \cR}}

\def\cQb{{\cal \Qb}}
\def\cRb{{\cal \Rb}}
\def\cWb{{\cal \Wb}}

\def\fd {{\rm N}}
\def\afd {{\overline{\rm N}}}

\def \II {I\hspace{-.1em}I\hspace{.1em}}
\def \IIA {\mbox{\II A\hspace{.2em}}}
\def \IIB {\mbox{\II B\hspace{.2em}}}
\def \gs {g^s}
\def \ls {\lambda^s}

\def \I {{\cal I}}
\def \qs {q\hspace{-.53em}/\hspace{.15em}}
\def \ks {k\hspace{-.53em}/\hspace{.15em}}
\def \YM {{\mbox{\tiny YM}}}
\def \gym {g_{\YM}}

\def \Lc {\L_c}
\def\IR{\relax{\rm I\kern-.18em R}}

\author{Chong-Sun Chu, Shou-Huang Dai and Douglas J Smith \\  
Centre for Particle Theory
and Department of Mathematics, 
Durham University, Durham, DH1 3LE, UK \\
E-mail:  
\email{chong-sun.chu@durham.ac.uk}, 
\email{shou-huang.dai@durham.ac.uk}, 
\email{douglas.smith@durham.ac.uk} }

\title {AdS/CFT Duality for Non-Anticommutative  
Supersymmetric Gauge Theory}

\abstract{We construct type IIB supergravity duals of non-anticommutative deformed
${\cal N} = 4$ SU($N$) gauge theories. We consider in particular deformations
preserving ${\cal N} = (1,0)$ and ${\cal N} = (1/2,0)$ supersymmetry. 
Such theories can
be realised on $N$ D3-branes in specific self-dual 5-form backgrounds. We show
that the required 5-form field strengths can be produced by configurations of
intersecting D3-branes and we are then able to construct the supergravity solutions
in the near-horizon limit. We consider some consequences of this duality, in
particular showing that the gravity duals predict that the dimensions of a
subset of BPS operators are not modified by the deformation.}
 
\preprint{}
\keywords{Non-Commutative Geometry,  AdS/CFT Correspondence, D-Branes}

\begin{document}

%\maketitle

\section{Introduction}

The AdS/CFT correspondence \cite{mal,gkp,witten,rev} is an explicit
realization of the holographic principle \cite{hol1,hol2}. It is the best
understood example of a gauge/gravity duality and offers promising
opportunities to tackle and understand some of the most difficult
problems in theoretical high energy physics such as the quantum properties
of spacetime and the confinement problem in gauge theory. 

%c1 Both these problems are well known to be notoriously difficult.

%c1 %c2
Owing to different motivations, various generalizations of the dualities
have been considered. We note in particular the ones \cite{MR,HI} for
noncommutative supersymmetric Yang-Mills and \cite{LM} where the duality
is characterized by a very interesting extended action of $SL(2,Z)$ 
\cite{dhk,CK}. 
Both of these deformations of the original AdS/CFT duality are motivated by  
having a deformed $*$-product
on the field theory side. In the first case, the spacetime is deformed
by a Moyal product induced by a constant NSNS B-field which lives on the
worldvolume of the D3-branes, while in the second case, the product
between fields carrying different $U(1)$ charges is deformed due to a
nontrivial twist in the TsT-transformation. In this regard, it is
natural and of interest to construct a gauge/gravity duality for the
non-anticommutative supersymmetric gauge theory \cite{seiberg} 
%c6
\cite{Klemm:01} where the fermionic
coordinates of the superspace are deformed with a non-anticommutative
$*$-product\footnote{
%c6
Non-anticommutative ${\cal N}=4$ super Yang-Mills theory can also be realized through deforming the constraint equations defined on the Euclidean superspace $\IR^{4|16}$\cite{Saemann:04}.}. 
The goal of this paper is to construct the
%c2 AdS/CFT 
gauge/gravity duality
for the non-anticommutative deformed $\cN=4$ supersymmetric Yang-Mills
theory. 
 
Non-anticommutative supersymmetric theories preserve a chiral
fraction of the supersymmetries. 
%c1 
That this is possible is because these theories
are defined in Euclidean space and the left and right chiral sectors are
not related by a complex conjugation. Due to their different supersymmetric
structure, a priori these theories could have quite different
quantum properties from their undeformed cousins. 
%c1 
Although power counting non-renormalizable, nevertheless they
are renormalizable \cite{renorm1,renorm2} 
basically because in a Feynman diagram computation, the
Hermitian conjugate partners which would be needed to generate divergent
counter terms are missing.
%c1 
Moreover due to the
existence of a superspace formulation, non-renormalization theorems exist
as usual. In addition
to having a very interesting mechanism of supersymmetry breaking,
non-anticommutative supersymmetric theories also possess interesting
non-perturbative properties \cite{nonpert}. 

In the original Maldacena AdS/CFT correspondence, the amount of
preserved supersymmetry is maximal. Since holography is believed to be a
generic property of quantum gravity, it is interesting to
understand how gauge/gravity duality works in a less or non-supersymmetric
setting,
%c4
especially when the supersymmetry is preserved in a non-standard
manner.
This is another motivation for our goal.

Non-anticommutativity in string theory \cite{ov,bgn,bs} 
was first discovered by Ooguri
and Vafa \cite{ov}, who observed that a self-dual graviphoton field
strength $C_{\m\n}$ induces a deformation in the fermionic part of the
4-dimensional superspace. Seiberg proposed another type of
deformation which imposes commutativity in the chiral coordinates
\cite{seiberg}. The deformation keeps $\cN=1/2$ supersymmetry 
in the case of simple supersymmetry, or
more specifically, $\cN=(1/2,0)$   of the
original $\cN=(1/2,1/2)$ supersymmetry \footnote{In Euclidean space, the 
Grassmannian-odd coordinates $\th^\a$ and $\thb^{\da}$ are not related
by complex conjugation, therefore it is more convenient to denote the 
simple $\cN=1$ supersymmetry as $\cN = (1/2,1/2)$. For the more general 
extended case, one can have
$\cN=(n/2,m/2)$ supersymmetry where $n,m$ are the number of left and right 
chiral spinorial 
supersymmetry generators.}. 
The deformed superspace has algebra  
\bea 
&& \{ \th^\a, \th^\b\} = C^{\a\b}, \label{CR-1}\\
&& \{ \thb^{\da}, \thb^{\db}\} = \{ \th^{\a}, \thb^{\db}\} =0, \\
&& [y^\m, y^\n] = [y^\m, \th^\a] = [y^\m, \thb^\da] =0,
\eea
where $y^\m = x^\m + i \th^\a \s^\m_{\a\da} \thb^\da$. The deformation 
is described by the constant $C^{\a\b}= C_{\m\n}(\s^{\m\n})^{\a\b}$. 

The generalization to the extended supersymmetry is immediate. 
For the $\cN=2= (1,1)$ case, the deformation generalizes to  
\be \label{CR-2}
\{ \th^{\a i}, \th^{\b j} \}= C^{\a\b ij}, \quad i=1,2, 
\ee
with all the other (anti)commutative relations remaining undeformed. The 
deformation parameter $C^{\a\b ij}$ obviously 
satisfies $C^{\a\b ij} = C^{\b\a ji}$. 
The deformation parameter can be decomposed into irreducible parts 
\cite{olaf1,FS}
\be \label{C2}
C^{\a\b ij} = \e^{\a\b} \e^{ij} I+ C^{(\a\b) (ij)} .
\ee
The deformation described by the first term is called the singlet deformation. 
It 
preserves Euclidean $SO(4)$ invariance and $SU(2)$
$R$-symmetry, and breaks $\cN=(1,1)$ supersymmetry down to $\cN=(1,0)$.
The singlet deformation can be obtained from string theory in a constant 
RR scalar 
background \cite{olaf-QS1}. 
The deformation described by the second term in \eq{C2} is 
called the non-singlet deformation. It retains $\cN=(1,0)$
supersymmetry for generic $C^{\a\b ij}$. 
However for particular deformation parameters
such that
\be \label{Cb}
C^{\a\b ij} = C^{\a \b} b^{ij} 
\ee
and with $\det b =0$,
the preserved supersymmetry is enhanced to 
$\cN=(1,1/2)$ \cite{olaf1}. 
The non-singlet deformation can be obtained from string theory in a constant
RR 5-form background 
%c1
\cite{olaf-QS1,ito-string2,ito-string4}. 
$\cN=4$ lightcone superspace could be defined. However
non-anticommutative deformations of it 
have not been considered. 

The non-anticommutative deformation \eq{Cb} can be obtained
from string theory in a particular RR 5-form background of the form
%c1 \cite{ito-string2,ito-string4} 
\be \label{RR5-1}
F_{\m\n a b c} = f_{\m\n} g_{abc},
\ee
where $\m,\n =0,1,2,3$ denote the 4-dimensional indices and 
$a,b,c,= 4, \cdots, 9$ are the indices of  the transverse space.
This 5-form is  self-dual both in the 4-spacetime
directions and in the transverse 6 dimensions
\be \label{RR5-2}
f_{\m \n} = \frac{1}{2!} \e_{\m\n\r\s} f_{\r\s}, \quad
g_{abc} = \frac{-i}{3!} \e_{abcdef} g_{def}.
\ee
In other words, the RR 5-form has the non-vanishing components $F_{\m\n abc}$ 
and satisfies the ``double self-duality'' condition
\bea \label{double-sd}
F_{\m\n abc} = \frac{1}{2!} \e_{\m\n\r\l} F_{\r\l abc} , \nn\\
F_{\m\n abc} =\frac{-i}{3!}  \e_{abcdef} F_{\m\n def}.
\eea
Note that $g_{abc}$ and hence the RR 5-form is necessarily complex since we are 
dealing with Euclidean signature. 

The action for the non-anticommutative SYM theory can be obtained using 
the superspace construction. For the deformed $\cN=(1/2,1/2)$ superspace, see
\cite{seiberg} for pure SYM and \cite{ito-sym1} for SYM theory with matter.
For the deformed $\cN=(1,1)$ case, one may use 
harmonic superspace. See 
\cite{ito-general,ito-QS-susy,olaf-QS1,olaf-QS2} for the case with
singlet deformation, and
\cite{ito-general,ito-QNS-susy,ito-QNS-ext,olaf-QNS} for the non-singlet
deformation. It can also be obtained from string theory as the worldvolume
action of D3-branes. More specifically,  the pure SYM action with
$\cN=(1/2,0)$ or  $\cN=(1,0)$ supersymmetry
can be obtained as the worldvolume action of D3-branes in a orbifold 
with constant graviphoton background \cite{billo,ito-string2}. 
Deformations of the $\cN=4$ SYM action with $\cN=(1/2,0)$ or $\cN=(1,0)$ 
supersymmetry can be obtained as the worldvolume action of D3-branes 
in  a specific configuration of RR 5-form flux \cite{ito-string4}. 
In this paper we will be interested in
constructing a gauge/gravity duality for the
%c1
non-anticommutative 
deformed $\cN=4$ SYM theories with $\cN=(1/2,0)$ 
and $\cN=(1,0)$ supersymmetry.

The paper is organized as follows. In section 2, we present the
construction of the 
%c1 
non-anticommutative  deformed $\cN=4$ SYM
theories with $\cN=(1/2,0)$ and $\cN=(1,0)$ supersymmetry as
the worldvolume action of D3-branes with a particular configuration of RR
5-form flux. In section 3, we realize the configurations as
intersecting brane systems and obtain the supergravity 
%c3 
duals of these non-anticommutative  deformed $\cN=4$ gauge theories. In
section 4, we focus on the theory with  $\cN=(1,0)$ supersymmetry and
analyse the duality. In particular we perform a standard
bulk-to-boundary analysis to extract  the two point
correlation function for the field theory. 
We find that the deformation modifies only the  overall coefficient,
but leaves the form of the two-point function unchanged. This 
implies that there exists a sector of BPS operators whose dimensions
are unmodified by the deformation.

\section{The Non-Anticommutative Deformations of the $\cN=4$ SYM}

\subsection{D3-brane realization of  $\cN=4$ SYM}

Consider $N$ D3-branes in the
0123-directions. The Lorentz group $SO(10)$ is decomposed into $SO(4)
\times SO(6)$ and the spin fields can be decomposed as $(S^\a
S^A,S_{\da} S_A)$ and $(\St^\b \St^B,\St_{\db} \St_B)$ where $S^\a,
\St^\a$ and $S_\da, \St_\da$ ($\a,\da =1,2$) are four dimensional
Weyl spinors and $S_A, \St_A$ and $S^A, \St^A$ ($A=1,2,3,4$) are 
six-dimensional Weyl spinors. 
The presence of a constant RR background can be described using the RR
vertex operator. In the $(-\frac{1}{2}, -\frac{1}{2})$ picture, 
the RR vertex operator  takes the form
\be
V_\cF = (2\pi \a)^{3/2} \St^T \cC \cF  S  e^{-\phi/2} e^{-\phit/2}  
\ee
where $\cF := \sum_p F_{\m_1 \cdots \m_{p+1}} \G^{\m_1 \cdots \m_{p+1}}/p!$ and 
$\cC$ is the charge conjugation matrix 
\footnote{
$\cF$ and $F_{\m_1 \cdots \m_{p+1}}$ are of dimension $[L]^{-2}$ here.
This is different from the normal dimension of $[L]^{-1}$ 
for the RR gauge field strength in supergravity theory
\be
S = (2 \k_{10}^2)^{-1} \int d^{10} x \sqrt{g} ( R -\frac{1}{2} F^2).
\ee 
% since $[R]= [L]^{-2}$.
}. 
Decomposing the spinor indices with respect to  $SO(4) \times SO(6)$, we have
\be
V_\cF = (2\pi \a)^{3/2} \cF^{\a\b AB} S_\a S_A \St_\b \St_B e^{-\phi/2}
e^{-\phit/2} + \cdots,
\ee
where 
$\cdots$ denotes contributions from the components of $\cF$ other than
$\cF^{\a\b A B}$. 
Due to their different tensor structure, these components 
will not be relevant for our discussion. 

To obtain the tensor structure of the deformation relations \eq{CR-1} or 
\eq{CR-2}, it
is necessary to consider a configuration of RR fields such that the only
non-vanishing components are the symmetric ones $\cF^{(\a\b) (AB) }$. 
Here $(\a \b)$, $(AB)$ represent symmetrization of
the indices. 
This can be achieved by turning on the
RR 5-form configuration \eq{RR5-1}, \eq{RR5-2}. As a result
\be \label{F-fg}
\cF^{\a\b AB} = f_{\m\n} g_{abc} (\s^{\m\n})^{\a\b} (\S^{abc})^{AB},
\ee
where
\be
\s^{\m\n} := \frac{1}{4} (\s^\m \sb^\n - \s^\n \sb^\m), \quad
\S^{abc} := \S^{[a} \Sb^b \S^{c]}
\ee
and they are self-dual:
\be
\s^{\m\n} = \frac{1}{2!} \e^{\m\n\r\l} \s^{\r\l}, 
\quad
\S^{abc} = \frac{i}{3!} \e^{abcdef} \S^{def},
\ee
in which $\s_\mu$, $\sb_\mu$ and $\S^a$, $\Sb^a$ are gamma 
matrices for the 4-dimensional and the 6-dimensional spaces; 
one example of the basis 
is given in Appendix.

The quantization of the string worldsheet
coupled to the RR-fields leads to the non-anticommutative relations 
\eq{CR-1} and \eq{CR-2} \cite{ov,bgn,bs,olaf-QS1}, and one expects 
the D3-brane worldvolume action to possess supersymmetry that is 
carried by the  deformed superspace.  
The SYM action on the worldvolume of the  D3-branes in the presence of
a constant RR 5-form flux $\cF^{(\a\b)(A B)}$ satisfying the double 
self-dual condition \eq{double-sd} was  
computed in \cite{ito-string4} using string perturbation theory. 
The deformation is determined by the parameters
\be \label{C-F}
C^{\a\b AB} := (2 \pi \a')^{3/2} \cF^{\a\b A B},
\ee
which are kept fixed in the $\a' \to 0$ limit. 
The action was computed up to the first order in $\cF$. 
The worldvolume action possess $\cN=(1/2,0)$ supersymmetry when $\cF^{\a\b AB}$ 
is of rank one in the $(A,B)$-space
%c1 
\cite{ito-string4}. 
When it is of rank two in the $(A,B)$-space, one expects the worldvolume
action to have $\cN=(1,0)$ supersymmetry.
An alternative
way to construct the deformed supersymmetric action is to use deformed
$\cN=(1,1)$ harmonic superspace.

%c1
%c1 Note that that the range of the indices $A, B$ determines
%c1 the portion of the supersymmetry that is deformed.
%c1 It turns out simpler to construct a deformation of the extended $\cN=(1,1)$ 
%c1 superspace,  so let us consider it first. 

\subsection{Non-anticommutative SYM with $\cN=(1,0)$ supersymmetry}

\subsubsection{RR-flux configuration for $\cN=(1,0)$ supersymmetry}

To introduce  a deformation to the $\cN=(1,1)$ superspace, the RR-5 form
$\cF^{\a\b AB}$ should be non-vanishing only for a   
$2 \times 2$ sub-block of the indices for $A, B$.
This can be achieved with the following configuration of RR 5-form:
\be \label{RR5-2by2}
\begin{array}{ccccccccc}
&& F_{01456} &=& -i F_{01789} &=& F_{23456} &=& -i F_{23789} =c, \\
&& F_{01786} &=& -i F_{01459} &=& F_{23786} &=& -i F_{23459} =c, 
\end{array}
\ee
where 
\be
c: = F_{01456}
\ee
is a  constant.
The first or the second line of \eq{RR5-2by2} is respectively 
a minimal configuration which satisfies \eq{RR5-1}, \eq{RR5-2}.
By having this particular combination of these minimal configurations,
$\cF$ is given by
\bea \label{F-M}
\cF^{\a\b A B} &=& 24 c (\s^{01} + \s^{23})^{\a\b} 
(\S^{456} + i\S^{459})^{AB} \nn \\
&=& 24 i c (\t^3)^{\a\b} M^{AB}
\eea
and 
\be  \label{M-def}
M := \S^{456} + i\S^{459}.
\ee

To proceed further, one needs 
%c2 
an explicit representation of the $\S$-matrices. For example, one can 
identify the $\S$-matrices here 
with the canonical choice of 
$\S$-matrices \eq{sigma6} given in the appendix. 
For example if we take
\be \label{id1}
\S^{6,9,4,5,7,8} = \big(\S^{4,5,6,7,8,9}\big)_{\rm appendix},
\ee
then 
\be \label{M0}
M = 2 i \left(
\begin{array}{cc}
\t^1 & 0 \\
0 & 0 
\end{array}
\right) :=M_0,
\ee 
which is of rank 2. 
We remark that a 
different identification of the $\S$-matrices gives a different $M$ that is
related to \eq{M0} by a 
bi-unitary transformation 
\be \label{M-UV}
 M = \Vt M_0 V,
\ee 
where $\Vt = V^T$, $V$ are unitary. The transformation 
is bi-unitary since in general $\Vt V \neq 1$. 
For example, for  the identification
\be
\S^{6,9,4,5,7,8} = \big(\S^{4,8,6,7,5,9}\big)_{\rm appendix},
\ee
we have
\be 
M = V^T \; 2 i \left(
\begin{array}{cc}
\t^1 & 0 \\
0 & 0 
\end{array}
\right) V, 
\ee 
with
\be
V = \frac{1}{\sqrt{2}} 
\left(
\begin{array}{cc}
\t^1 &  -\t^2 \\
-\t^2 & \t^1 
\end{array}
\right).
\ee
It is
\be
V^T V= -V V^T =  -i \left(
\begin{array}{cc}
0 & \t^3 \\
\t^3 & 0 
\end{array}
\right) \neq 1
\ee 
and so the transformation \eq{M-UV} is not unitary, but a bi-unitary
one.
We note that the vertex operator with $M$ given by \eq{M-UV} is
equivalent to the one with $M = M_0$ under a 
change of basis for the spin field $S_A, \St_A$
\be
%c2 inverse taken out
S   \to V S, \quad
\St  \to   V \St.
\ee
Thus we have shown that by turning on the constant RR 5-form field
\eq{RR5-2by2}, $\cF^{\a \b A B}$ takes the factorized form \eq{Cb} with
$\det b \neq 0$, i.e. it is of rank 2 with respect to the $A,B$ indices. 
 
\subsubsection{Harmonic superspace and non-anticommutative SYM}

Given the non-anticommutative relation \eq{CR-2}, 
the deformed $\cN=4$ action on the worldvolume of the D3-branes can  
be obtained as a supersymmetric action for the gauge superfield 
and hypermultiplet superfield of deformed harmonic
superspace. Let us first give a brief introduction to 
harmonic superspace \cite{hss}. For a
comprehensive review, we refer the reader to \cite{hss-book}. 

Let $(x^\m,
\th_i^\a, \thb^{\da i})$ be the coordinates of the $\cN=(1,1)$ superspace,
where $\m=0,1,2,3$ are the spacetime indices, $\a,\da =1,2$ spinor
indices and $i=1,2$ are the  indices of the $SU(2)$ R-symmetry. 
The harmonic superspace is supplemented by the harmonic variables
$u_i^\pm$ which form an $SU(2)$ matrix:
\bea \label{u-def}
&& u^{i+} u_i^- =  1, \quad u^{+i} u_i^+ = u^{-i} u_i^- =0, \nn\\
%c2 change to tilde
&& \widetilde{u^{+i}} = u_i^-. 
\eea
%c2 
Here $~\widetilde{}{}~$ is the standard conjugation acting on the harmonic
superspace. Its action  on the other coordinates of the superspace is
\be
\tilde{x^\m_A} =x^\m_A, \qquad 
\widetilde{\th^{\pm \a}} = \ve_{\a\b}\th^{\pm \b}, \qquad  
\widetilde{\thb^{\pm \da}} = \ve_{\da\db} \thb^{\pm \db},
\ee
%c3 
where 
\bea
&& \th^\pm_\a := u_i^\pm \th^i_\a, \quad \thb^\pm_\da := u_i^\pm \thb^i_\da,
\nn\\
%c6
&& x_A^\m = x^\m - i (\th^+ \s^\m \thb^- + \th^- \s^\m \thb^+)
\eea
is the analytic basis of the harmonic superspace.
And as a result we have the condition on the parameters deforming
the $\cN=(1,1)$ superspace:  
\be
\widetilde{C^{\a\b ij}} = C_{\a\b ij }.
\ee

Using these variables, one can introduce the harmonic projection of the
supercovariant derivatives
\be
D_\a^\pm = u_i^\pm D^i_\a, \quad \bar{D}_\a^\pm = u_i^\pm \bar{D}^i_\a. 
\ee
Instead of chiral superfields, one considers analytic superfields in harmonic 
superspace. They satisfy $D_\a^+ \Phi = \bar{D}_\da^+ \Phi =0$.
The supercharges and supercovariant derivatives take the form
\bea
Q^+_\a &=& \frac{\del}{\del \th^{-\a}} - 2i \s^\m_{\a\da} \thb^{+\da}
\frac{\del}{\del x^\m_A}, \quad Q^-_\a = -\frac{\del}{\del \th^{+\a}}
\nn\\
\Qb^+_\da &=& \frac{\del}{\del \thb^{-\da}} + 2i \th^{+\a}\s^\m_{\a\da}
\frac{\del}{\del x^\m_A}, \quad \Qb^-_\da = -\frac{\del}{\del
\thb^{+\da}},\nn\\
D^+_\a &=& \frac{\del}{\del \th^{-\a}}, \quad 
D^-_\a = - \frac{\del}{\del \th^{+\a}} + 2i \s^\m_{\a\da} \thb^{-\da}
\frac{\del}{\del x^\m_A},\nn\\
\Db^+_\da &=&  \frac{\del}{\del \thb^{-\da}},\quad
\Db^-_\da = - \frac{\del}{\del \thb^{+\da}}
- 2i \th^{-\a}  \s^\m_{\a\da}  \frac{\del}{\del x^\m_A},
\eea
in terms of which the condition for the analytic superfield can be solved 
easily and is of the form 
\be
\Phi =\Phi(x_A^\m, \th^+, \thb^+, u).
\ee
One can expand the analytic superfield in $\th$ and obtain a finite
expansion with coefficients being functions of $x_A$ and $u$. 
Each $\th$-component can be further expanded in terms of symmetrized products 
of $u^+$ and $u^-$. This second expansion is infinite
and so each analytic superfield contains an infinite number of component
fields.

It is convenient to 
introduce covariant derivatives with respect to $u^\pm$
compatible with the defining relations \eq{u-def}
\bea
 D^{++}&=& u^{+i} \frac{\del}{\del u^{-i}}, \quad D^{--}= u^{-i}
\frac{\del}{\del u^{+i}},\nn\\
D^{0} &=&  u^{+i} \frac{\del}{\del u^{+i}} - u^{-i} \frac{\del}{\del
u^{-i}}.
\eea
The operator $D^0$ measures the $U(1)$ charges of the harmonics $u^\pm$.
A function of charge $q$ satisfies
\be
D^0 \phi^{(q)} =q \phi^{(q)}
\ee
and admits the expansion (here $q \geq 0$; 
for $q<0$ there is an analogous formula)
\be
\phi^{(q)}(u) = \sum_{n=0}^\infty \phi^{(i_1 \cdots i_{q+n} j_1 \cdots
j_n)} \; u^+_{(i_1} \cdots u^+_{i_{q+n}}u^-_{j_1} \cdots u^-_{j_n
)}.
\ee
The coefficients
$\phi^{i_1 \cdots j_m}$ are irreducible $SU(2)$ tensors with isospin
$(n+m)/2$.

The  non-anticommutative $\cN=(1,1)$ superspace has the deformed relation
\be \label{CR}
\{ \th^{\a i}, \th^{\b j} \}= C^{\a\b ij}, \quad i=1,2. 
\ee
The deformation is equivalent to  a $*$-product 
\be
(f * g)(\th) = f(\th) \exp \Big( 
-\frac{1}{2}\stackrel{\longleftarrow}{\frac{\del}{\del \th_i^\a}} 
C^{\a\b}_{ij} \stackrel{\longrightarrow}{\frac{\del}{\del \th_j^\b}}
\Big) g(\th).
\ee
In the harmonic superspace approach, the $\cN=2$ gauge multiplet is
described by a charge 2 analytic superfield $V^{++} = V^{++ M} T^M$
where $T^M$ are the Lie algebra generators. We will consider $U(N)$ in
this paper. Under the deformed $U(N)$ gauge group, the gauge multiplet
transforms as
\be
\d_\L V^{++} = - D^{++} \L + i [V^{++} \stackrel{*}{,} \L]
\ee
where $\L$ is an analytic superfield parameter. The action of $\cN=2$
SYM is given by \cite{zupnik}, \cite{olaf1}
\be
S_{V} = \frac{1}{2} \sum_{n=2}^\infty \frac{(-i)^n}{n}\tr
\int d^{12} z 
du_1 \cdots d u_n \frac{ V^{++}(z,u_1)* V^{++}(z,u_2) *\cdots *V^{++}(z,u_n)}
{(u_1^+ u_2^+) (u_2^+ u_3^+) \cdots (u_n^+ u_1^+)},
\ee
where $z=(x, \th_i^\a, \thb_i^\da)$. 

As for the hypermultiplet, it can be described either by a complex
analytic superfield $q^+$ with $U(1)$ charge +1 or by a real analytic
neutral superfield $\o$. These descriptions are known to be related to
each other via a duality \cite{hss-book} and one can restrict to
either description. 
%c4 
Similar consideration applies in the deformed case.
 To construct the
deformed $\cN=4$ SYM,  let us consider $q^+$ in the adjoint representation. 
The coupling of $q^+$ to the
$\cN=2$ gauge multiplet is given by
\be
S_q = - \int d \z  du  \Tr \; \qb^+ *(D^{++} + i [V^{++}
\stackrel{*}{,}q^+]).
% \big[ (D^{++} + i V^{++})* \o\big]^2  
\ee
where 
\be
d \z  := d^4 x_A d^4 \th^-.
\ee

The $\cN=4$ SYM theory can be written down in terms of these $\cN=2$
superfields. In fact, by using an $\cN=2$ gauge multiplet and an $\cN=2$
hypermultiplet $q^+$ in the adjoint representation, the $\cN=4$ action 
can be written as
\be \label{S-sf}
S_{SYM} = S_V + S_q.
\ee
For a generic non-singlet deformation \eq{Cb} with generic $b$ such that 
$\det b \neq 0$, the theory has $\cN=(1,0)$ supersymmetry. 

The action
\eq{S-sf} is written in terms of $\cN=(1,1)$ superfields. To rewrite it in
terms of component fields, one needs to substitute the
%c2  
component expansion of the
superfield and carry out the integrals
in $\th$ and $u$. As we remarked above, the component expansion of an
analytic superfield contains an infinite number of component
fields and auxiliary fields. However many of these can be gauged away. For
example, one can utilize the infinite degrees of freedom 
present in the analytic gauge parameter $\L$ to eliminate all the 
auxiliary fields in the gauge 
superfield  $V^{++}$. In the Wess-Zumino gauge, $V^{++}_{WZ}$ has only a
finite number of physical components \cite{olaf1}. For the hypermultiplet
superfield $q^+$, the auxiliary fields can be eliminated from the
action using the classical equation of motion for $q^+$. 
%c1 See
We refer the reader to
\cite{olaf1} for the case of a $U(1)$ gauge group. The generalization to
$U(N)$ is straightforward. 

Although the resulting component action is manifest in supersymmetry, the
gauge transformations of the component fields are typically obscured and
become non-canonical. This was first observed in \cite{seiberg} for the
deformed $\cN =(1/2,1/2)$ superspace. To obtain component fields which have
canonical gauge transformations, one must perform a field
redefinition. The redefined component fields admit canonical gauge
transformations, but their supersymmetry transformations are deformed. 
This can be worked out explicitly and fully in the deformed $\cN=(1/2,1/2)$
case. However this becomes much more complicated for the deformed
$\cN=(1,1)$ case \cite{ito-general}, \cite{ito-QS-susy}, \cite{olaf-QS1},
\cite{ito-QNS-susy}. Both the field redefinition and the deformed
supersymmetry transformations involve infinite series expansions in the
deformation parameter. Therefore although there is in principle no
difficulty to write down the deformed action explicitly, the
procedure is rather involved and we will not carry out its evaluation
here.

\subsection{Non-anticommutative SYM with $\cN=(1/2,0)$ supersymmetry}

To construct a deformation of the $\cN=4$ SYM theory with $\cN=(1/2,0)$
supersymmetry, one can first write the $\cN=4$ theory in  terms of
$\cN=(1/2,1/2)$ superfields and then introduces non-anticommutative deformation 
to the $\cN=(1/2,1/2)$ superspace. In this case, the RR-5 form
$\cF^{\a\b AB}$ should be non-vanishing only for a   
$1 \times 1$ sub-block of the indices for $A, B$.

This can be achieved by turning on further components in addition to
those in \eq{RR5-2by2} which has the effect of further reducing the rank of
$\cF^{\a\b A B}$. Up to equivalence, the appropriate RR 5-form
configuration is 
\be \label{RR5-1by1}
\begin{array}{ccccccccccc}
&& F_{01456} &=& -i F_{01789} &=& F_{23456} &=& -i F_{23789} &=& c, \\
&& F_{01786} &=& -i F_{01459} &=& F_{23786} &=& -i F_{23459} &=& c, \\
&& F_{01476} &=& i  F_{01589} &=& F_{23476} &=&  i F_{23589} &=& i c, \\
&& F_{01586} &=& i  F_{01479} &=& F_{23586} &=&  i F_{23479} &=& -i c, \\  
\end{array}
\ee
where 
\be
c: = F_{01456}
\ee
is a constant.
The matrix $\cF$ takes the form
\be \label{F-M1}
\cF^{\a\b A B} 
= 24 i c (\t^3)^{\a\b} M^{AB},
\ee
where in this case 
%c5
$M^{AB}$ is given by
\be  \label{M-def-1}
M := \S^{456} + i\S^{459} + i(\S^{476} + i\S^{479}) .
\ee

Without loss of generality, we take the same identification of $\S$-matrices 
as %c2
in \eq{id1} to obtain
\be 
M = 4 i U^T M_0 U,
%c := U^T M_0  U ,
\ee
where
\be
M_0= \left(
\begin{array}{cccc}
1 & 0 & 0 & 0 \\
0 & 0 & 0 & 0 \\
0 & 0 & 0 & 0 \\
0 & 0 & 0 & 0 
\end{array}
\right)
\quad
\mbox{and}
%c5
\quad  
U = \frac{1}{\sqrt{2}} 
\left(
\begin{array}{cccc}
1 &  1 & 0 & 0 \\
-1 & 1 & 0& 0 \\
0 & 0 & 1 & 0 \\
0 & 0 & 0 & 1 
\end{array}
\right).
\ee
As before, the presence of $U$ is a matter of choice of basis 
and can be absorbed by a transformation of the spin fields. Therefore the 
configuration \eq{RR5-1by1} of the RR 5-form flux gives rise to a 
deformation that is  governed by \eq{F-M1} with $M$ of rank 1, and
corresponds to a non-anticommutative deformation of the $\cN=(1/2,1/2)$
superspace.

As mentioned above, the non-anticommutative deformed SYM theory can be
obtained easily using the deformed $\cN=1$ superspace. The theory admits
$\cN=(1/2,0)$ supersymmetry. It is interesting to note that the
additional terms in the action which deform the theory have an
interpretation as the Chern-Simons couplings of the D3-brane to a
certain constant RR 5-form background \cite{im}.

\section{The Supergravity Solution}
\label{SUGRASoln}
  
It is easy to check that the constant RR 5-form field strength
\eq{RR5-2by2} does not generate any energy-momentum tensor in flat
Euclidean space:
\be
%c4
T_{MN} = F_{M M_1 M_2 M_3 M_4} F_N{}^{M_1 M_2 M_3 M_4} 
- \frac{1}{10} g_{MN} F^2 =0 .
\ee
However this is no longer the case once one takes into account
the backreaction of the $N$ D3-branes, which turns the flat spacetime
to $AdS_5 \times S^5$. Our goal now is to construct the supergravity
solution which would give rise to the components \eq{RR5-2by2} for the
RR 5-form field on the worldvolume of the $N$ D3-branes. Moreover, as a
deformation, the solution should reduce back to the original $AdS_5
\times S^5$ background when the deformation is turned off.

\subsection{Supergravity dual for the $\cN=(1,0)$ case}
\label{sugra}

In order to obtain the desired configuration \eq{RR5-2by2} 
of the  RR 5-form flux, we
consider the following configuration of intersecting D3-branes,
\be \label{brane-sys}
\begin{array} {c|ccccccccccc}
D3_1    & (&0& 1& 2& 3&  &  & &  &  &)\\
D3_2    & (&0& 1&  &  & 4& 5& &  &  &)\\
D3_{2'} & (&0& 1&  &  &  &  & &7 &8 &)\\
D3_{3}  & (& &  & 2& 3& 4& 5& &  &  &)\\
D3_{3'} & (& &  & 2& 3&  &  & &7 &8 &)
\end{array}.
\ee
Here $D3_1$ denotes the original $N$ D3-branes; and we have introduced
four additional sets of D3-branes.
Let us check supersymmetry. In type IIB string theory, 
the two supersymmetries $\ve_1, \ve_2$ are of the same  chirality.
The set of $N$ D3-branes imposes the condition
\be \label{susy-1}
\G^{0123} \ve_1 = \ve_2.
\ee
This condition relates the two supersymmetries and hence
reduces the supersymmetry by one half. Now introduce the other 4 sets of branes 
$D3_2, D3_{2'}$, $D3_3, D3_{3'}$. This imposes additionally the
conditions
\be\label{susy-2}
\G^{0145} \ve_1 = \ve_2 ,\quad \G^{0178} \ve_1 = \ve_2, \quad 
\G^{2345} \ve_1 = \ve_2, \quad \G^{2378} \ve_1 = \ve_2.
\ee
The 4 conditions in \eq{susy-2} are not all independent. In fact there are only
3 independent equations in \eq{susy-2} and \eq{susy-2}  is equivalent to 
the following system:
\be \label{susy-3}
\G^{2378} \ve_1 = \ve_2, \quad \G^{0123} \ve_1 = - \ve_1, \quad
\G^{4578} \ve_1 = -\ve_1.
\ee
Together with \eq{susy-1}, we see that generically 
our set of intersecting branes preserves
1/16  of the type IIB supersymmetry, i.e. 2 supersymmetries are preserved.
%c2 This is the generic situation. 
However in the near horizon limit of the 
$N$ D3-branes,  the condition \eq{susy-1}
is lifted and all the 32 supersymmetries are preserved. Therefore, in this
limit,  we only have the conditions \eq{susy-3}. The first of these conditions
gives $\ve_2$ once $\ve_1$ is solved. The second and the third conditions 
in \eq{susy-3} impose 2 conditions on $\ve_1$ which means 4
supersymmetries are preserved. Moreover the 4 supersymmetries are chiral
both in the 4-dimensional and in the 6-dimensional sense. Hence we can
denote the preserved supersymmetries by
\be
\ve^{\a A}, \qquad\quad  \a=1,2; \quad A=1,2.
\ee
This matches precisely with the preserved $\cN=(1,0)$ 
supersymmetries in the non-anticommutative SYM theory.

The metric of our intersecting branes system is given by
\bea
ds^2 &=& \sqrt{\frac{H_3 H_{3'}}{H_1 H_2 H_{2'}}} (dx_0^2 + dx_1^2)
+ \sqrt{\frac{H_2 H_{2'}}{H_1 H_3 H_{3'}}} (dx_2^2 + dx_3^2)
+\sqrt{\frac{H_1 H_{2'} H_{3'}}{H_2 H_{3}}} (dx_4^2 + dx_5^2)\nn\\
&& + \sqrt{\frac{H_1 H_2 H_{3}}{H_{2'} H_{3'}}} (dx_7^2 + dx_8^2)
+ \sqrt{H_1 H_2 H_3 H_{2'} H_{3'}} (dx_6^2 + dx_9^2)
\eea
and the RR 5-form is
\be
F= F_0 + F_1,
\ee
where
\bea
%c2 change sign
F_0 &:=& d(\frac{1}{H_1} ) dx^{0123} +  \mbox{dual}, \label{F0}\\
F_1 &:=& d(\frac{1}{H_2} )  dx^{0145} +d(\frac{1}{H_{2'}} )  dx^{0178}
+d(\frac{1}{H_{3}} ) dx^{2345}  +d(\frac{1}{H_{3'}} )  dx^{2378} 
+   \mbox{dual}. \;\;\;\quad\; \label{F1}
\eea
$F_0$ is the RR 5-form sourced by the original set of $N$ D3-branes, and
$F_1$ is sourced by the additional sets of branes.

%c1
In order that no components of the RR 5-form other than those 
that are present in  
\eq{RR5-2by2} are activated, we choose the harmonic functions 
$H_2, H_{2'}, H_3, H_{3'}$ to be functions of $x_6$ and $x_9$ only. 
Moreover to produce the complex structure of the RR 5-form in the equation
\eq{RR5-2by2}, it is necessary that $H_2$ and $H_{2'}$ depend on
$x_6, x_9$ in a particular way:
\be \label{holo}
H_2 = H_2 (z), \quad H_{2'} = H_{2'} (z), \quad
H_3 = H_3 (z), \quad H_{3'} = H_{3'} (z)
\ee
where
\be
z = x_6 +i x_9
\ee
is a complex variable. In other word, the branes $D3_2, D3_{2'},
D3_3, D3_{3'}$ are smeared and have effectively a single transverse direction.

%c3 The equations of motion for the system are:
The equations of motion for this system of partially localised intersecting
branes are given by the curved space Laplace equations \cite{IB}
\be\label{H1}
(H_2 H_{3} \del_i^2 + H_{2'} H_{3'} \del_m^2 + \del_a^2) H_1 =0,
\ee
\be\label{H2}
\del_a (\frac{H_2^2}{H_{3'}^2} \del_a (\frac{1}{H_2}) ) =0,
\ee
\be\label{H2p}
\del_a (\frac{H_{2'}^2}{H_{3}^2} \del_a (\frac{1}{H_{2'}}) ) =0,
\ee
\be\label{H3}
\del_a (\frac{H_{3}^2}{H_{2'}^2} \del_a (\frac{1}{H_{3}}) )=0,
\ee
\be\label{H3p}
\del_a (\frac{H_{3'}^2}{H_2^2} \del_a (\frac{1}{H_{3'}}) )=0,
\ee
where we have used $i=4,5$ to denote the indices in the $x_4,x_5$
directions and $\del_i^2 := \del_4^2 +\del_5^2$ is the 2-dimensional flat 
Laplacian. Similarly $\del_m^2 :=\del_7^2 +\del_8^2$ and 
$\del_a^2 := \del_6^2+ \del_9^2$. Due to \eq{holo}, 
the equations \eq{H2} -\eq{H3p} are satisfied immediately.
Since the branes $D3_2, D3_{2'},
D3_3, D3_{3'}$ are smeared and have effectively a single transverse
direction, the charge associated with them is well defined only if 
$F_1$ as given by 
\eq{F1} is well-defined at $|z| =\infty$. Moreover we would like to
reproduce the components of the RR flux precisely at the worldvolume of
the set of $N$ D3-branes, i.e. at $x_4 = x_5= x_6= x_7= x_8= x_9 =0$.
We obtain
%c1
the unique solution
\be
H_2 = H_{2'} =H_3= H_{3'} = \frac{1}{1+ c z}.
\ee
In this case, the RR 5-form field strength \eq{F1} is actually constant
and equal to \eq{RR5-2by2} everywhere.

Finally the equation for $H_1$ reduces to 
\be \label{de-H1}
\big( \del_i^2 +  \del_m^2 + \frac{1}{H_2^2} \del_a^2\big) H_1 =0 .
\ee
Naively one may try to treat $c$ as a small parameter and solve
\eq{de-H1} by solving the differential equation perturbatively.   
This is messy however. A much simpler way to solve \eq{de-H1} in closed form 
is due to the following observation. We first 
rewrite the Laplacian $\del_a^2 = 4 \del_z
\del_{\zb}$ where $\zb := x_6- i x_9$ and introduce
\be \label{map1}
w: = \int H_2^2 dz = \frac{z}{1+c z},
\ee
where we have chosen the integration constant such that $w=0$ when
$z=0$. 
%c2
In terms of $w$ and $\zb$,
the equation \eq{de-H1} can be rewritten as
\be \label{de-H1-new}
\big( \del_i^2 +  \del_m^2 + 4 \del_w \del_{\zb}) H_1 =0 .
\ee
Except for the fact that $w$ is not the complex conjugate of $\zb$, 
this is formally the same Laplace equation as in the undeformed
$AdS_5\times S^5$ case. This fact allows us to solve \eq{de-H1-new}
easily. We obtain
\be \label{H1-soln}
H_1 = 1+ \frac{R^4}{\rho^4}, 
\ee
where 
\be 
\frac{R^4}{\a'^2}:= 4 \pi g_s N  = \l 
\ee
and 
\be \label{rho2}
\rho^2:= x_i^2 + x_m^2 + w \zb 
\ee
It is 
\be
\rho^2 = x_i^2 + x_m^2 + \frac{z\zb}{1+c z}
=  x_i^2 + x_m^2 + \frac{w \wb}{1-\cb \wb}
\ee
Formally the solution $H_1$ takes the same form as the undeformed one.
That this is true is because the 
%c1 
differential algebra involved does not care about the
complex structure. 
%Substitute \eq{H1-soln}
Using the above results, 
%c2 and definitions, 
the metric becomes
\be
ds^2  = \frac{1}{\sqrt{H_1}} (dx_0^2 + dx_1^2 +dx_2^2 + dx_3^2) + 
\sqrt{H_1}\Big(dx_4^2 + dx_5^2 + dx_7^2 + dx_8^2 + \frac{dz d\zb}{(1+cz)^2} 
\Big).  
%c2 \nn\\
%c2 &=& \frac{1}{\sqrt{H_1}} (dx_0^2 + dx_1^2 +dx_2^2 + dx_3^2) + 
%c2 \sqrt{H_1}\Big(dx_4^2 + dx_5^2 + dx_7^2 + dx_8^2 + \frac{dw d\wb}{(1-\cb
%c2   \wb)^2} \Big). \qquad \quad \eea
\ee
%c2 
It is clear that the singularity at $z= -1/c$ is infinitely
far away and so the supergravity  background is regular. We also remark
that although the metric is invariant under $SO(4)$ rotations in $x^\m$,
$\m=0,1,2,3$, the Euclidean Lorentzian symmetry 
$SO(4)$ is broken by the RR 5-form. This agrees with  the field theory.

Summarizing, our proposal is that the non-anticommutative SYM theory with
deformation parameter \eq{C-F}, \eq{RR5-2by2} 
is dual to the near horizon limit of the supergravity background given by 
the intersecting brane system
\eq{brane-sys}. The near horizon limit is taken with 
\be
\xt^a := x^a/\a'
%c \quad \mbox{and} \quad  \a' x^\m := \xt^\m 
\ee  
fixed in the $\a' \to 0$ limit. 
Introducing
\be
U:= \rho/\a'
\ee
and also scaling $c$ such that 
\be \label{c-scale}
\ct : = \a' c = \a' F_{01456} % \quad \mbox{is fixed in the $\a' \to 0$ limit.}
\ee
is fixed in the $\a' \to 0$ limit, we obtain the near horizon metric
%c2 
\bea
\label{NHMetric}
\frac{ds^2}{\a'} &=& \frac{U^2}{\sqrt{\l}} d x_\m^2 
+\frac{\sqrt{\l}}{U^2} \Big(d\xt_4^2 + d\xt_5^2 + d\xt_7^2 + d\xt_8^2 
+ \frac{d \zt d\bar{\zt}}{(1+\ct \zt )^2} \Big),
\eea
where $\zt:=z/\a'$, 
%c2 \be \wt = \frac{\zt}{1+ \ct \zt} \ee
%c2
and
\be
U^2 =  \xt_i^2 + \xt_m^2 + \frac{\zt \bar{\zt}}{1+ \ct \zt}.
\ee
The RR 5-form is
\be
F= F_0 + F_1,
\ee
where
\be
%c2 sign 
\frac{F_0}{\a'^2} =   d(\frac{U^4}{\l}) d x^{0123} +  \mbox{dual} 
\ee
and 
\bea 
\frac{F_1}{\a'^2} = &&  \ct (d x^0 dx^1 d\xt^4 d\xt^5 d\xt^6  
+ i d x^0 dx^1 d\xt^7 d\xt^8 d\xt^9 
+ d x^2 dx^3 d\xt^4 d\xt^5 d\xt^6
+  i d x^2 dx^3 d\xt^7 d\xt^8 d\xt^9 \nn \\
&&+  id x^0 dx^1 d\xt^4 d\xt^5 d\xt^9
+  d x^0 dx^1 d\xt^7 d\xt^8 d\xt^6
+  i d x^2 dx^3 d\xt^4 d\xt^5 d\xt^9
+   d x^2 dx^3 d\xt^7 d\xt^8 d\xt^6    )\nn\\
&& + \mbox{dual}. \qquad\;
\eea
Note that 
$F_1$ is well defined in the same limit \eq{c-scale} where the metric has 
a well-defined limit. Note also that the conditions 
\eq{c-scale} and \eq{C-F} 
are indeed the same due to different normalization. 

We remark that the RR-flux is necessarily complex in order to
generate the non-anticommutative deformation. This is also reflected in the
complexification of the metric through \eq{replace}.
We also remark that the effect of turning on the deformation \eq{RR5-2by2} in 
the gauge theory is  a simple  replacement
\be \label{replace}
d z d \zb \to dw d \zb
\ee
in the metric of the supergravity dual. It is remarkable that the effects of
non-anticommutativity can be summarized nicely in such a compact form through
a simple change of variables.
%c1
We will further comment on this in the discussion section.

\subsection{Supergravity dual for the $\cN=(1/2,0)$ case}
\label{sugra1}

To obtain the configuration \eq{RR5-1by1} 
of the  RR 5-form flux, we
consider the following configuration of intersecting D3-branes,
\be \label{brane-sys-1}
\begin{array} {c|ccccccccccc}
D3_1    & (& 0& 1& 2& 3&  &  & &  &  &)\\
D3_2    & (& 0& 1&  &  & 4& 5& &  &  &)\\
D3_{2'} & (& 0& 1&  &  &  &  & &7 &8 &)\\
%c6
D3_3    & (&  &  & 2& 3& 4& 5& &  &  &)\\
D3_{3'} & (&  &  & 2& 3&  &  & &7 &8 &)\\
D3_4    & (& 0& 1& & & 4 &  & & 7 &  &)\\
D3_{4'} & (& 0& 1&  &  & & 5& &  & 8 &)\\
D3_5    & (&  & & 2 & 3 & 4 &  & &7 & &)\\
D3_{5'} & (&  & & 2 & 3 &  & 5 & & &8 &)
\end{array}.
\ee
Here $D3_1$ denotes the original $N$ D3-branes; and we have introduced
eight additional sets of D3-branes.
The checking of supersymmetry is similar as before. We have from 
the original $N$ D3-branes the condition 
\be \label{susy-1-1}
\G^{0123} \ve_1 = \ve_2,
\ee
and the conditions
\be\label{susy-2a-1}
\G^{0145} \ve_1 = \ve_2 ,\quad \G^{0178} \ve_1 = \ve_2, \quad 
\G^{2345} \ve_1 = \ve_2, \quad \G^{2378} \ve_1 = \ve_2,
\ee
\be\label{susy-2b-1}
\G^{0147} \ve_1 = \ve_2 ,\quad \G^{0185} \ve_1 = \ve_2, \quad 
\G^{2347} \ve_1 = \ve_2, \quad \G^{2385} \ve_1 = \ve_2,
\ee
from the additional sets of branes. In the near horizon limit, this is
equivalent to 
\be \label{susy-3a-1}
\G^{2378} \ve_1 = \ve_2, \quad \G^{0123} \ve_1 = - \ve_1, \quad
\G^{4578} \ve_1 = -\ve_1
\ee
and
\be \label{susy-3b-1}
\G^{48} \ve_1 = \ve_1.
\ee
The presence of the additional projection condition \eq{susy-3b-1} 
reduces further the
unbroken supersymmetry to a single two-component chiral spinor and we
can denote it as 
\be
\ve^{\a}, \qquad \a=1,2.
\ee
Therefore, in the near horizon limit,  
the intersecting branes configuration preserves
1/16  of the type IIB supersymmetry, i.e. 2 supersymmetries.
This matches precisely with the  $\cN=(1/2,0)$ 
supersymmetries in the non-anticommutative SYM theory.

The metric of our intersecting branes system is given by
\bea
ds^2 &=& 
\sqrt{\frac{H_3 H_{3'} H_5 H_{5'}}{H_1 H_2 H_{2'} H_4 H_{4'}}} 
(dx_0^2 + dx_1^2)
+ \sqrt{\frac{H_2 H_{2'}  H_4 H_{4'}}{H_1 H_3 H_{3'}  H_5 H_{5'}}} 
(dx_2^2 + dx_3^2) \nn\\
&+& \sqrt{\frac{H_1 H_{2'} H_{3'}  H_{4'} H_{5'}}{H_2 H_{3}  H_4 H_{5} }} 
dx_4^2 
+\sqrt{\frac{H_1 H_{2'} H_{3'}  H_{4} H_{5}}{H_2 H_{3}  H_{4'} H_{5'} }} 
dx_5^2\nn\\
&+& \sqrt{\frac{H_1 H_{2} H_{3}  H_{4'} H_{5'}}{H_{2'} H_{3'}  H_4 H_{5}
}}  dx_7^2 
+ \sqrt{\frac{H_1 H_{2} H_{3}  H_{4} H_{5}}{H_{2'} H_{3'}  H_{4'} H_{5'} }} 
dx_8^2 \nn\\
&+& \sqrt{H_1 H_2 H_3 H_4 H_5 H_{2'} H_{3'} H_{4'} H_{5'} } (dx_6^2 + dx_9^2)
\eea
and the RR 5-form is
\be
F= F_0 + F_1,
\ee
where
\bea
F_0 &:=& d(\frac{1}{H_1} ) dx^{0123} +  \mbox{dual}, \label{F0-1}\\
F_1 &:=& d(\frac{1}{H_2} )  dx^{0145} +d(\frac{1}{H_{2'}} )  dx^{0178} 
+d(\frac{1}{H_{3}} ) dx^{2345}  +d(\frac{1}{H_{3'}} )  dx^{2378} \label{F1-1}
\\
&& +d(\frac{1}{H_4} )  dx^{0147} +d(\frac{1}{H_{4'}} ) dx^{0158}  
+d(\frac{1}{H_{5}} )  dx^{2347}
+d(\frac{1}{H_{5'}} )  dx^{2358}
+   \mbox{dual}. \nn
\eea
$F_0$ is the RR 5-form sourced by the original set of $N$ D3-branes, and
$F_1$ is sourced by the additional sets of branes. As before, we need to
choose the functions $H_2$, $H_{2'}$, $H_3$, $H_{3'}$, $H_4, H_{4'}, H_5, H_{5'}$ 
to be functions of $z=x_6+ i x_9$ only.

The equations of motion for the system are:
\be\label{H1-1}
\Big(H_2 H_{3} (H_4 H_5 \del_4^2 + H_{4'} H_{5'} \del_5^2) 
+ H_{2'} H_{3'} (H_4 H_5 \del_7^2 + H_{4'} H_{5'} \del_8^2)
+\del_a^2 \Big) H_1 =0, 
\ee
\bea 
\label{H2-1}
\del_a (\frac{H_4 H_{4'}}{H_5 H_{5'}} 
\frac{H_2^2}{H_{3'}^2} \del_a (\frac{1}{H_2}) ) &=&0,\\
\label{H2p-1}
\del_a (\frac{H_4 H_{4'}}{H_5 H_{5'}}
\frac{H_{2'}^2}{H_{3}^2} \del_a (\frac{1}{H_{2'}}) ) &=&0,\\
\label{H3-1}
\del_a (\frac{H_5 H_{5'}}{H_4 H_{4'}}
\frac{H_{3}^2}{H_{2'}^2} \del_a (\frac{1}{H_{3}}) )&=&0,\\
\label{H3p-1}
\del_a (\frac{H_5 H_{5'}}{H_4 H_{4'}}
\frac{H_{3'}^2}{H_2^2} \del_a (\frac{1}{H_{3'}}) )&=&0,\\
\label{H4-1}
\del_a (\frac{H_2 H_{2'}}{H_3 H_{3'}} 
\frac{H_4^2}{H_{5'}^2} \del_a (\frac{1}{H_4}) ) &=&0,\\
\label{H4p-1}
\del_a (\frac{H_2 H_{2'}}{H_3 H_{3'}} 
\frac{H_{4'}^2}{H_{5}^2} \del_a (\frac{1}{H_{4'}}) ) &=&0,\\ 
\label{H5-1}
\del_a (\frac{H_3 H_{3'}}{H_2 H_{2'}} 
\frac{H_5^2}{H_{4'}^2} \del_a (\frac{1}{H_5}) ) &=&0,\\
\label{H5p-1}
\del_a (\frac{H_3 H_{3'}}{H_2 H_{2'}} 
\frac{H_{5'}^2}{H_{4}^2} \del_a (\frac{1}{H_{5'}}) ) &=&0,
\eea
where $a=6,9$. The equations \eq{H2-1} - \eq{H5p-1} are satisfied
immediately.
As before, since the additional sets of branes are smeared 
and have effectively a single transverse
direction, the charge associated with them is well defined only if 
$F_1$ as given by 
\eq{F1-1} is well-defined at $|z| =\infty$. Moreover we want to
reproduce the components \eq{RR5-1by1} 
of the RR flux precisely at the worldvolume of
the set of $N$ D3-branes. We obtain the 
%c1 
unique solution
\bea
&& H_2 = H_{2'} =H_3= H_{3'} = \frac{1}{1+ c z}, \nn\\
&& H_4 = H_{5} = \frac{1} {1+ic z} \nn\\
&& H_{4'}= H_{5'} = \frac{1}{1 -i c z},
\eea
The RR 5-form field strength \eq{F1-1} is  constant
and equal to \eq{RR5-1by1} everywhere. 

Finally, the equation for $H_1$ reduces to 
\be \label{de-H1-1}
\big( A \del_i^2 +  \frac{1}{A} \del_m^2 + \frac{1}{H_2^2 H_4 H_{4'} } 
\del_a^2\big) H_1 =0 ,
\ee
where
%c2
$i=4,7$, $m=5,8$, $a=6,9$ here, and the function $A$ is defined by
\be \label{A}
A := \frac{H_4}{H_{4'}} = \frac{1-icz}{1+icz}. 
\ee
The differential equation \eq{de-H1-1} can be solved as
follows. Introduce the change of variable
\be \label{wz}
w := \int H_2^2 H_4 H_{4'} dz = 
\frac{z}{2(1+cz)}
 +\frac{1}{2c} \ln(1+cz) - \frac{1}{4c} \ln(1+c^2 z^2),
\ee
the equation \eq{de-H1-1} for $H_1$ becomes
\be \label{de-H1-2}
\big( A \del_i^2 +  \frac{1}{A} \del_m^2 + 4 \del_w \del_{\zb} \big) H_1 =0 .
\ee
This can be solved with the ansatz
\be
H_1 = 1+ \frac{R^4}{\rho^4},
\ee
where
\be
\rho^2 = B_1(w) x_i^2 + B_2(w) x_m^2 + C(w) \zb.
\ee
The equation \eq{de-H1-2} is satisfied if the following conditions on the
coefficient functions $B_1(w), B_2(w)$ and $C(w)$ hold:
\be \label{ode1}
C'  - 3C \frac{B_1'}{B_1} +\frac{B_2}{A} -2 A B_1 =0, 
\ee
\be \label{ode2}
C' - 3C \frac{B_2'}{B_2} +B_1 A -2 \frac{B_2}{A} =0,
\ee
\be \label{ode3}
C' = \frac{1}{2}(B_1 A + \frac{B_2}{A}),
\ee
%c2 
where $'$ here refers to differentiation with respect to $w$.
It is easy to obtain from these equations
\bea
(B_1 B_2)' =0, \label{ode4}\\
(\frac{C}{B_1})' = A,\label{ode5}\\
(\frac{C}{B_2})' =\frac{1}{A}. \label{ode6}
\eea
By rescaling $\rho$, we can set the integration constant of
\eq{ode4} to be 1 and we obtain 
\be
B_1 = 1/ B_2.
\ee
The equations \eq{ode5} and \eq{ode6} then give
\be
 C B_1 = \int  \frac{1}{A} dw, \quad \frac{C}{B_1} = \int A dw ,
\ee
and hence
\be \label{sol_B}
B_1(w(z)) = \frac{1}{B_2(w(z))} = \sqrt{\frac{N(z)}{D(z)}} 
\ee
\be \label{sol_C}
C(w(z)) = \sqrt{N(z) D(z)}
\ee
where 
\be \label{sol_ND}
N(z) := \int \frac{1}{A} dw , \quad D(z):= \int A dw. 
\ee
Substituting the definition \eq{A} for $A(z)$ and recalling \eq{wz}, we have
%c5
\be \label{sol_N}
N(z) = \frac{1}{4c} \Big[
(1-i) \ln\frac{(1+cz)^2}{1+c^2 z^2}  + 2(1+i) \tan^{-1}(cz)
  -2(1+i) \frac{c^2 z^2}{(1+cz)(i+cz)} \Big],  
\ee
\be \label{sol_D}
D(z) = \frac{1}{4c} \Big[
(1+i) \ln\frac{(1+cz)^2}{1+c^2 z^2}  + 2(1-i) \tan^{-1}(cz)
  -2(1-i) \frac{c^2 z^2}{(1+cz)(-i+cz)} \Big].   
\ee
We also record the small $c$ expansions
\bea
B_1 &=& 1+ i c z + O(c^2 z^2), \\
B_2 &=& 1- i c z + O(c^2 z^2), \\
C &=& z (1-cz + O(c^2 z^2)), 
\eea
where it is clear that the solution reduces back to the undeformed one
$B_1= B_2 =1$, $C=z$ when $c \to 0$. 

The near horizon limit is given as before by taking
\be
\xt^a := x^a/\a', \quad \mbox{and} \quad U:= \rho/\a'
\ee  
fixed in the $\a' \to 0$ limit.  
We also scale $c$   such that 
\be 
\ct : = \a' c, 
\ee
is fixed in the $\a' \to 0$ limit.  
We obtain the near horizon metric
\bea
%c2 \label{NHMetric-1}
\frac{ds^2}{\a'} &=& \frac{U^2}{\sqrt{\l}} d x_\m^2 
+\frac{\sqrt{\l}}{U^2} \Big(\frac{1}{\At} (d\xt_4^2 + d\xt_5^2) 
+\At (d\xt_7^2 + d\xt_8^2)  + 
\frac{1}{(1+\ct \zt)^2(1+\ct^2\zt^2)}d\zt d\bar{\zt} \Big),
%c2 
\nn
\eea
where we have defined 
\be
\zt:=z/\a', 
\ee
The function $\At$ is given by
\be
\At:= \frac{1-i \ct \zt  }{1+i \ct \zt  } 
\ee
and 
\be
U^2 = B_1 \xt_i^2 + B_2 \xt_m^2 
+ \frac{C}{\a'} \bar{\zt},
\ee
where, with a slight abuse of notation, 
the coefficients $B_1, B_2$ and $C$ are obtained from \eq{sol_B}-\eq{sol_D} by
replacing $c z$ with $\ct \zt$ everywhere.
The RR 5-form is given by 
${F_0}/{\a'^2} =   d(U^4) d x^{0123} /\l +  \mbox{dual}$,  
together with the constant components \eq{RR5-1by1}.
This supergravity solution is the dual for the non-anticommutative deformed
$\cN=4$ SYM with $\cN=(1/2,0)$ supersymmetry. 

\section{Some Consequences of the Correspondence}

In this section we will consider the effect of the non-anticommutative
deformation on the field theory, as predicted by the supergravity dual. 
We will concentrate on the duality with $\cN=(1,0)$ supersymmetry 
since in this case the supergravity background 
as constructed in section \ref{sugra} 
%c1 
is slightly simpler.  
In particular we will analyse the anomalous dimensions of the
field theory operators.

As already noted in section~\ref{sugra} the metric is formally identical
to the $AdS_5 \times S^5$ metric of the undeformed theory, subject to the
replacement of $z$ with $w = z/(1+cz)$. However,
there is one subtlety which we must first address. Since we have deformed the
field theory in a non-Hermitian way, the dual supergravity solution has been
modified by a complex deformation. The resulting effect is the replacement of
$z$ with $w$, but for $c \ne 0$, 
%c $\overline{w} \ne \overline{z}$. 
$w \ne z$.
We therefore must be
careful when interpreting this geometry. In particular, we need to identify
the conformal boundary of the spacetime to relate bulk and boundary fields.
Since we are viewing this theory as a deformation of the $\cN =4$ theory,
we will use the standard notion of the boundary as $r \rightarrow \infty$
where $r^2 = x^i x_i + x^m x_m + |z|^2$. For $c \ne 0$ this differs from the
complex quantity, $\rho^2 = x^i x_i + x^m x_m + w\overline{z}$ which naturally
appears in many quantities. However, generically $\rho$ diverges in the limit
$r \rightarrow \infty$.

We will now consider the correspondence between bulk scalar fields and field
theory operators. From the metric in equation~(\ref{NHMetric}) a scalar field
$K$ with mass $m$ satisfies the Laplace equation which implies that
\be
\left( \frac{\lambda}{\rho^4}\partial_{\mu}^2 + \partial_i^2 + \partial_m^2 
+ 4\partial_w\partial_{\overline{z}} 
- m^2\frac{\sqrt{\lambda}}{\rho^2} \right) K = 0
\ee
As for the undeformed case, solutions of this equation which are independent
of the ``5-sphere'' are given by
\be
K = \frac{\xi^{\Delta}}{(x^{\mu}x_{\mu} + \xi^2)^{\Delta}}
\ee
where $\xi = 1/\rho$ and $\Delta = 2 + \sqrt{4+m^2}$. We will now see that,
despite the distinction between $\rho$ and $r$, these states are dual to field
theory operators with scaling dimension $\Delta$. Therefore there is a class
of field theory operators whose spectrum is not deformed. Note however that there
are two possible ways in which the spectrum of operators can be deformed. There
are the other solutions to the ten-dimensional Laplace equation which have a
dependence on the ``5-sphere''. Since this is deformed, the resulting spectrum
of 5-dimensional masses will be changed. Also, the ten-dimensional spectrum of
the full string theory is likely to depend on the deformation, giving a
dependence of $m^2$ on $c$ for string theory states.

We will now use the above solution $K(\xi,x^{\mu})$ to give the 5-dimensional
bulk to boundary propagator for these scalars and calculate the two-point
function of the dual field theory operators using standard techniques. So, a
boundary field $\phi_0(x^\mu)$ is a source for the bulk field configuration
\be
\phi(\xi, x^{\mu}) = \int \mathrm{d}^4x' 
\frac{\xi^\Delta}{(|x-x'|^2 + \xi^2)^{\Delta}}\phi_0(x')
\ee

To calculate the two-point function we need to consider the dependence of the
action for the bulk scalar field on its ``boundary values'' $\phi_0(x)$. 
This is
\bea
I & = & \int\mathrm{d}^4x\mathrm{d}r\mathrm{d}^5\Omega \frac{1}{2}\sqrt{g}
\left( \partial_M\phi\partial_N\phi g^{MN} + m^2 \phi^2 \right) \\
& = & \frac{1}{2}\int\mathrm{d}^4x\mathrm{d}^5\Omega 
\left( \sqrt{g} \phi g^{rN} \partial_N\phi \right)_{r \rightarrow \infty}
\eea
where we have used integration by parts and the equation of motion for $\phi$
to perform the integral over $r$.

We can now use the above relation between $\phi$ and $\phi_0$, together with
the following standard polar parametrisation of the coordinates
\bea
x^6 + ix^9 & = & r\cos\alpha \mathrm{e}^{i\phi_1} \\
x^4 + ix^5 & = & r\sin\alpha\cos\theta \mathrm{e}^{i\phi_2} \\
x^7 + ix^8 & = & r\sin\alpha\sin\theta \mathrm{e}^{i\phi_3}
\eea
to write the action explicitly in terms of the boundary sources $\phi_0$. Note
that for fixed angles, $\xi \rightarrow 0$ as $r \rightarrow \infty$.
Explicitly, we find
\bea
\sqrt{g}  &=&  r^5\xi^2 H_2\sin^3\alpha\cos\alpha \sin\theta \cos\theta \\
\phi(\xi \rightarrow 0, x) & =&  \xi^{4-\Delta}\phi_0(x) \\
g^{rN} \partial_N\phi  &=&
-\frac{\Delta}{H_2^2}\left( (1 + (H_2^2-1)\sin^2\alpha)\partial_r\rho 
+ \frac{1}{r}(H_2^2-1)\sin\alpha\cos\alpha\partial_{\alpha}\rho
\right)\times \nn\\
&&  \times \int \mathrm{d}^4x' \frac{\phi_0(x')}{|x-x'|^{2\Delta}}
\eea
where in the last result we have kept only the leading order terms as
$\xi \rightarrow 0$. Putting everything together we find the expression for the
two-point function of the operator dual to the scalar field
\bea
\langle \cO(x) \cO(x') \rangle 
& = & \frac{\delta I}{\delta\phi_0(x) \delta\phi_0(x')} \nonumber \\
 & = & -\frac{\Delta}{2} \frac{1}{|x-x'|^{2\Delta}} 
\int \mathrm{d}^5\Omega \sin^3\alpha\cos\alpha \sin\theta \cos\theta 
\frac{r^5}{\rho^5 H_2 } \times \nonumber \\
 &   & \,\, \times \left( (1 + (H_2^2-1)\sin^2\alpha)\partial_r\rho 
+ \frac{1}{r}(H_2^2-1)\sin\alpha\cos\alpha\partial_{\alpha}\rho \right) \;\; \\
 & = & \frac{C}{|x-x'|^{2\Delta}}
\eea
where the constant $C$ is given by the integral in the previous line which is
to be evaluated in the limit $r \rightarrow \infty$. Actually evaluating the
integral is not straightforward since the large $r$ behaviour is different for
$\alpha = 0$ and $\alpha \ne 0$, as can easily be seen by considering the
explicit expression for 
$\rho^2 = r^2\sin^2\alpha 
+ r^2\cos^2\alpha/(1+cr\cos\alpha \mathrm{e}^{i\phi_1})$. 
Nevertheless the final result clearly indicates
that the anomalous dimension of the operator $\cO$ is given by $\Delta$. Hence,
for this class of operator, the only corrections to this dimension, as compared
to the undeformed theory, can come from the possible dependence of the bulk
mass $m$ on the deformation parameter $c$.

Similar results will follow for field theory operators dual to other bulk
fields. Due to the nature of the deformation, we expect that the spectrum of
BPS states is simply a subset of the BPS states in the undeformed geometry.
This would then correspond to the prediction that the scaling dimensions of
the chiral operators in the field theory are the same as in the $\cN = 4$
theory, but that the rest of the theory 
%c is likely to
will be deformed.

\section{Discussions}

In this paper we have constructed the supergravity duals for the
non-anticommutative deformed $\cN=4$ supersymmetric Yang-Mills theory 
with $\cN=(1,0)$ and $\cN=(1/2,0)$ supersymmetries. The supergravity
solution consists of a metric which is a complex deformation of the
$AdS_5 \times S^5$ metric, 
and a RR 5-form fields with complex constant components. The fact that
the metric is non-dilatonic suggests that the field theory coupling
is not renormalized. It will be interesting to check this explicitly.

Deformed by non-anticommutativity of the fermionic components of the
superspace, the  non-anticommutative field theory breaks supersymmetry 
in a novel non-traditional
way. Nevertheless it preserves many remarkable properties of the usual
supersymmetric field theories.  It will be interesting to analyse and
understand more this kind of  supersymmetric breaking from the supergravity
point of view. 

%c2 
The supergravity background dual to the non-anticommutative gauge
theory is complex. The imaginary nature of the RR 5-form is easy to
understand and is a direct
consequence of solving the self-duality condition in Euclidean space.
The imaginary nature of the metric is more obscure.
Although we have demonstrated that one can
nevertheless  extract
physical information such as the dimensions of operators in a more
or less the standard way using the bulk-to-boundary approach, it will
be good to have a 
%c4 better 
deeper understanding on the imaginary nature of the
metric.  We recall that the complexity of the metric (and the
flux background) is a direct reflection of the fact that the 
non-anticommutative field theory is non-Hermitian.
%c1 
With an analysis which is based on a reduction to the quantum mechanics, 
it has been
suggested \cite{crypto} that  non-anticommutative 
theory is unitary in a more general
sense \cite{PT}. This 
%c4 should have an explanation in terms of the
suggests something similar in the
dual supergravity description.
As a first step, it is natural to try to find
the corresponding phenomena in the supergravity side when a similar
reduction of degree of freedoms is performed.
In the mini-superspace approximation,
one can work out the canonical Hamiltonian. Due to
the presence of complex components in the metric,  the
Hamiltonian is not real.  We
conjecture that the Hamiltonian is pseudo-real in a sense similar to
its field theory counterpart. 
%c4 
%c4 If this is true, the complexity of the supergravity background will be
%c4 an apparent one and the supergravity background make perfect 
%c4 physical sense. 
This would provide a physical understanding of the nature of the
complexity of the supergravity background. 
We leave this interesting question for future
analysis.

%%%%%%%%%%%%%%%%%%%%%%%%%%%%%%%%%%%%%%%%%%%%%%%%%%%
%c2 \newpage
\section*{Acknowledgements}   
We thank Patrick Dorey, Pei-Ming Ho, Olaf Lechtenfeld and Katsushi Ito for
discussions. 
The work of CSC is supported in part by EPSRC and PPARC. The work of
DJS is supported by PPARC.  SHD's work 
is supported by a scholarship of the Ministry of Education, Taiwan.
 
\appendix

\section{Notation and Convention}

We denote by $\m=0,1,2,3$ the directions in the 4-dimensional Euclidean
worldvolume of the $D3$-branes, $a=4,5,6,7,8,9$ the transverse
directions, and
$\a,\da =1,2$ the spinor indices. We
follow the notation of \cite{wess}. In particular, spinor indices 
are raised and lowered by
the $\ve$-tensor with the convention  $\ve^{12} = -\ve_{12}
=1$.  

Decomposing $SO(10)$ into $SO(4) \times SO(6)$,
the ten-dimensional gamma matrices are given by
\be
\G_{(10)}^\m = \g^\m \otimes 1, \quad \G_{(10)}^a = \g^5 \otimes \G^a,
\ee
where
\be
\g^\m = \left(
\begin{array}{cc}
0 & \s^\m \\
\sb^\m & 0
\end{array}
\right),
\quad
\G^a = \left(
\begin{array}{cc}
0 & \S^a \\
\Sb^a & 0
\end{array}
\right).
\ee
Here the matrices $(\s^\m)_{\a\db}$ and $(\sb^\m)^{\da \b}$ are given by
\bea \label{sigma4}
\sigma_\mu &=& (i\tau^1,i\tau^2,i\tau^3, 1), \nn\\
\bar{\sigma}_\mu &=& (-i\tau^1,-i\tau^2,-i\tau^3,1),
\eea
where $\tau^i$ ($i=1,2,3$) are the Pauli matrices. They satisfy the
Clifford algebra
\be
\s_\m \sb_\n + \s_\n \sb_\m = 2 \d_{\m\n} {\bf 1}.
\ee
The Lorentz generators are defined by
\be
\sigma^{\mu \nu} = \frac{1}{4}
( \sigma^{\mu} \bar{\sigma}^{\nu} - \sigma^{\nu} \bar{\sigma}^{\mu}),
\quad
%c6
\bar{\sigma}_{\mu\nu}={\frac{1}{4}}(\bar{\sigma}_\mu\sigma_\nu-\bar{\sigma}_\nu
\sigma_\mu).
\ee
The matrices $(\s_{\m\n})_{\a\b}$,  $(\st_{\m\n})^{\da\db}$ are symmetric in the
spinor indices. 
Moreover $\s_{\m\n}$ is self-dual and  $\st_{\m\n}$ is anti self-dual
with respect to the $\m,\n$ indices.

The gamma matrices for six-dimensional part are
given by
\bea \label{sigma6} 
\Sigma^a &=&
\left( \eta^{3}, - i\bar{\eta}^{3}, \eta^{2}, - i\bar{\eta}^{2},
\eta^{1}, i\bar{\eta}^{1} \right),\nn\\
\bar{\Sigma}^a &=&
(-\eta^3, -i\bar{\eta}^3, -\eta^2,-i\bar{\eta}^2, -\eta^1, i\bar{\eta}^1), 
\eea
where $a=4,\cdots,9$. $\eta^a_{\mu\nu}$ and $\bar{\eta}^a_{\mu\nu}$ are
the
{}'t Hooft symbols, which are defined by
\be
\sigma_{\mu\nu}={\frac{i}{2}}\eta^a_{\mu\nu} \tau^a,
\quad 
\bar{\sigma}_{\mu\nu}={\frac{i}{2}}\bar{\eta}^a_{\mu\nu} \tau^a.
\ee
The matrices  \eq{sigma6} satisfy the Clifford algebra
\begin{equation}
(\Sigma^a)^{AB}(\bar{\Sigma}^b)_{BC}+(\Sigma^b)^{AB}(\bar{\Sigma}^a)_{BC}
=2\delta^{ab}\delta^A_C .
\end{equation}

The charge conjugation matrix is block diagonal  in this basis
\be
\cC = \cC_{(4)} \otimes \cC_{(6)},
\ee
where
\be
 \cC_{(4)} = \left(
\begin{array}{cc}
- \e^{\a \b}  & 0  \\
0 & - \e_{\da \db}
\end{array}
\right),\quad 
\cC_{(6)} = \left(
\begin{array}{cc}
0 & -i \d_A{}^B  \\
- i \d_B{}^A &0 
\end{array}
\right) . 
\ee

%%%%%     %%%%%     %%%%%     %%%%%     %%%%%     %%%%%     %%%%%     %%%%%    
%%%%%     %%%%%     %%%%%     %%%%%     %%%%%     %%%%%     %%%%%     %%%%%

\end{document}